\documentclass[12pt]{article}

\usepackage{amsthm,amsmath,amssymb}
\usepackage{mathrsfs}
\usepackage{times}
\usepackage{epsfig,amsmath}
\usepackage{subfigure}
\usepackage{graphicx}
\usepackage{color}
\usepackage{epstopdf}
\usepackage{cite}

\newenvironment{sciabstract}{%
\begin{quote} \bf}
{\end{quote}}

\newcounter{lastnote}

\title{Fast Correlated-Photon Imaging Enhanced by Deep Learning}

\author
{Zhan-Ming Li,$^{1,2}$ Shi-Bao Wu,$^{1,2}$ Jun Gao,$^{1,2}$ Heng Zhou,$^{1,2}$ Zeng-Quan Yan,$^{1,2}$\\ Ruo-Jing Ren,$^{1,2}$ Si-Yuan Yin,$^{1,2}$ Xian-Min Jin$^{1,2\ast}$\\
	\\
	\normalsize{$^1$Center for Integrated Quantum Information Technologies (IQIT), School of Physics  }\\
	\normalsize{and Astronomy and State Key Laboratory of Advanced Optical Communication Systems}\\
	\normalsize{and Networks, Shanghai Jiao Tong University, Shanghai 200240, China.}\\
	\normalsize{$^2$CAS Center for Excellence and Synergetic Innovation Center in Quantum Information}\\
	\normalsize{and Quantum Physics, University of Science and Technology of China, Hefei, 230026, China}\\
	\
	\normalsize{$^\ast$E-mail: xianmin.jin@sjtu.edu.cn}\\
}

\date{}

\begin{document} 
\baselineskip24pt

\maketitle

\begin{sciabstract}
Correlated photon pairs, carrying strong quantum correlations, have been harnessed to bring quantum advantages to various fields from biological imaging to range finding. Such inherent non-classical properties support extracting more valid signals to build photon-limited images even in low flux-level, where the shot noise becomes dominant as light source decreases to single-photon level. Optimization by numerical reconstruction algorithms is possible but require thousands of photon-sparse frames, thus unavailable in real time. Here, we present an experimental fast correlated-photon imaging enhanced by deep learning, showing an intelligent computational strategy to discover deeper structure in big data. Convolutional neural network is found being able to efficiently solve image inverse problems associated with strong shot noise and background noise (electronic noise, scattered light). Our results fill the key gap in incompatibility between imaging speed and image quality by pushing low-light imaging technique to the regime of real-time and single-photon level, opening up an avenue to deep leaning-enhanced quantum imaging for real-life applications.
\\
\end{sciabstract}

Correlated-photon imaging, relying on the inherent quantum correlations between entangled photon pairs, emerges as a novel technique bringing quantum enhancement to a large number of research fields \cite{jachura2015shot,Brida2009Measurement,chrapkiewicz2016hologram,sun2019mapping,qiu2019structured}. The direct imaging of non-classical correlations can reveal entanglement between position and momentum~\cite{howell2004realization,aspden2013epr} or entanglement among optical angular momentum modes~\cite{moreau2019imaging}.
The new imaging technique enabled by single photon-sensitive cameras can be used to test fundamental quantum physics~\cite{basden2003photon,jost1998spatial,zhang2009characterization,blanchet2010purely,toninelli2017sub,brida2010experimental,tang2018experimental,wang2019topological,xu2020scalable} and to improve the conventional imaging systems on spatial resolution and signal-to-noise ratio (SNR)~\cite{morris2015imaging,aidukas2019phase,aspden2015photon}.
 
Unfortunately, imaging system's performance at low-light flux is affected by shot noise due to the quantum nature of light. Intensified scientific complementary metal-oxide-semiconductor (I-sCMOS) cameras are able to capture single photons by virtue of image intensifier technology ~\cite{assmann2009higher,wiersig2009direct,assmann2010ultrafast,schwartz2013superresolution,edgar2012imaging}. In order to extract the single-photon signal from the noise, a reasonable threshold is set to binarize the data in each pixel, a signal over which we define successful registration of one photon~\cite{wen2006individual,lantz2008multi}. 

Reconstructing such photon-limited images can be converted to solve inverse problems~\cite{mccann2017convolutional,lucas2018using}. Numerical algorithms~\cite{harmany2011spiral,katsaggelos2007super,chen2014variational} can solve the inverse problems by treating Poisson statistics as the prior knowledge and performing complex iterative operations, such as least squares, maximum likelihood and convex optimization. Apparently, thousands frames of raw image data have to be collected to form a proper-precision statistical result, which therefore prevent the reconstruction implementing in real time. Machine learning, also known as ``end-to-end'' approach, can merge multiple stages into one single neural network~\cite{jain2009natural,burger2012image,xie2012image,wang2016non,chen2016trainable,zhang2017beyond}. It implies that it is possible to find a direct relationship between the original objective and measured ultra low-light images by learning from large data sets.

Here, we report an experimental deep learning of correlated photon imaging where the convolutional auto-encoder (CAE) is trained for extracting effective signals from various noises. As a result, deep learning algorithm shows superior performance over the numerical reconstruction algorithm in image restoration and superresolution at single-photon level, especially the ability to achieve high-contrast imaging in real time. Our results suggest emerging deep-learning-enhanced applications in quantum imaging and quantum information processing.

In general, the imaging measurements, $\mathit{y}$, is noisy compared with the original image, $\mathit{x}$, due to the imperfect imaging system, such as laser's Poisson properties, optical elements' limited size and camera's low quantum detection efficiency. However, a statistical model describing the forward imaging system becomes ill posed when the number of photons decreases. Thus regularization is often introduced into a designed numerical algorithm $\emph{R}_{\mathrm{reg}}$ to search solutions that match the prior knowledge about objects~\cite{harmany2011spiral}, 
$$\emph{R}_{\mathrm{reg}}(y)=\underset{\mathit{x}}{\operatorname{argmin}} {\mathcal L}\{\varPhi(\mathit{x}), y\}+\emph{h}(\mathit{x}) \eqno{(1)}$$
where \textbf{$\varPhi$} represents the Poisson forward model. \textbf{\emph{$\mathcal L$}} is loss function, an appropriate measure of error. $\emph{h}$ is a regularizer to control model complexity and reduce overfitting. The choice of the regularizer is often based on practical experience. The total variation (TV) regularization has been extensively studied and applied in the field of image denoising. It often converts the image denoising into a well-posed problem by introducing certain constraints, thus ensures the existence and uniqueness of the optimized image, so the method has the advantage of being less disturbed by noise.

As is shown in Fig.1(a), numerical algorithms usually consist of the following process: Pre-processing step rescales the measurement data to fit the inputs in Modeling step, where the reconstruction process is converted to a convex optimization program from Eq.(1). After sequential quadratic approximations, we get an optimized image corresponding to original object.

As an effective method for feature extraction and image denoising, deep learning has revolutionized our ability to use computers to perform challenging tasks. It provides another alternative, called the learning algorithm, as is shown in Fig.1(b). Original objects and their corresponding measurements, $\left\{\left(x_{n}, y_{n}\right)\right\}_{n=1}^{N}$, are fed into a neural network as inputs, the reconstruction algorithm, $\emph{R}_{\text {learn }}$, is trained for optimizing the following process 
$$\emph{R}_{\text {learn }}=\underset{\emph{R}_{\theta}, \theta}{\operatorname{argmin}} \sum_{n=1}^{N} \mathcal L \{x_{n},\emph{R}_{\theta}(y_{n})\}+\emph{h}(\theta) \eqno{(2)}$$ 
where \textbf{\emph{$\mathcal L$}} is loss function, $\emph{h}$ is a regularizer, and $\theta$ is possible parameters in the neural network.  Learning algorithm usually consists of two parts: the encoder maps the features of input images into the hidden layer space, and then the decoder restores these features to reconstruction images. The internal parameters are adjusted to minimize the loss function from Eq.(2) between measurements and original data. Once the learning step is complete, the neural network structure can serve as the optical inverse imaging model, and recover new images from their measurements in a straightforward fashion.

Our experimental arrangement is shown in Fig.2(a). The correlated photons are generated in a BBO crystal by spontaneous parametric down-conversion (SPDC) process. The idler photons are detected by a single-photon avalanche diode (SPAD) to trigger a multipixel I-sCMOS camera. The signal photons are reflected by a spatial light modulator (SLM) with image information and pass through a linear polarization analyzer, before being captured by the I-sCMOS camera. In order to ensure the synchronization of correlated photon pairs, we compensate the electronic delay time by adding a 10m fiber delay line in the camera arm. Wave plate and polarization beam splitter are employed to initiate and optimize signal photons by monitoring the fewest photons in the reflection path. We set the camera exposure time as 5s, the delay time as 19ns and the gate-width as 10ns, which corresponds to about 1.6 photon per pixel on average  (see Methods for details).  

The layout of the CAE network is schematically shown in Fig.2(b). We build the model with three convolutional layers, three maxpooling layers, three deconvolutional layers and three unpooling layers. The images are split into a training set and a test set containing 10521 and 1169 images from the Extended MNIST (EMNIST) handwritten letter database, respectively. Training letters are sequentially displayed on the amplitude-only SLM shown in Fig.2(a), and the corresponding noisy images are then captured by the I-sCMOS  camera. All input-output data are fed into the network to optimize the parameters of every layers, after it's complete, the performance of the CAE framework is tested using the test images (see Methods for details). 

We choose the ``photon'' letter samples to demonstrate the performance from different reconstruction algorithms, as is shown in Fig.3(a)-(d).  Direct measurements in the camera plane are very noisy compared with the original objects in low light level condition. TV regularization is suitable for optimizing photon-limited images, so background noise needs to be pre-filtered by above photon counting approach. When there is only one frame of data, this scheme has little effect on image reconstruction. In contrast, the CAE algorithm is very efficient in suppressing the shot noise and electronic noise. Besides, we plot an intensity map of one line of the ``n''  image, as is shown Fig.3(e). Sharp edges can be well reconstructed by the end-to-end method. The reasonable interpretation is that the CAE algorithm does not fully learn the forward imaging model, but rather learns how to suppress noise and also how to optimize the feature parameters of the training images. 

To quantify this excellent improvement, we define the image contrast as
$$C=\frac{I_{\max }-I_{\min }}{I_{\max }+I_{\min }} \eqno{(3)}$$
In the case of ultra weak signals, for TV-regularization, only a very faint image of the object is obtained with a contrast of 0.4, while the CAE algorithm recovers a better image with an enhanced contrast of 1. This result indicates that the CAE algorithm has an advantage in suppressing noise and fast image reconstruction.
 
Further, to verify the robustness of the learning algorithm, we prepare another data set containing 6690 handwritten digits downloaded from the MNIST database, in which 6021 images as a training set and 669 images as a test set. Furthermore, fewer correlated photons are illuminated on the samples and recorded by the camera with $ \sim $ 0.8 photons per pixel on average. We display ``1905''  digits in Fig.4(a)-(c). Fewer photons result in indistinguishable raw signal measurements. Interestingly, the reconstructed images still give good contrast. Thus, the CAE algorithm can protect the signals from noise damage and demonstrate strong robustness. 

Besides, in order to optimize the CAE structure, we constructed the CAE networks with 5 layers, 7 layers, and 9 layers. Fig.4(d) shows the cost changing with the number of epochs. After 1000 epochs, the mean-squared error (MSE) between the network output and the appearance of the handwritten digits drops down to 0.25 for 5 layers, and becomes steady. However, for 7 layers and 9 layers, the cost difference becomes smaller after 2000 epochs. In fact, using the least convolution layers is necessary to realize the best denoising results, since it can reduce the computational cost significantly and save huge computing resources.

We summarize the state-of-the-art single photon imaging experiments, as is shown in Fig.5. Imaging systems differ in various applications, leading a trade-off in visibility and the time spent on collecting data. Compared with the passive imaging schemes, active imaging enables higher contrast by high-precision nanosecond time gate filtering out noise from the signals. However, intensified CCD and CMOS architectures suffer from the low frame rate, as a result, the traditional reconstruction algorithms can enhance the image only by collecting thousands of sparse-photon frames, which waste a lot of time ~\cite{tang2018experimental,morris2015imaging,PhysRevLett.122.193903}. Another active imaging scheme is based on SPAD cameras capable of counting and time stamping single photons with picosecond time resolution. Single-pixel scanning imaging~\cite{kirmani2014first} with an excellent visibility acquires a megapixel scan in approximately 20 minutes. While SPAD array imaging~\cite{shin2016photon} achieves great improvement, but still requires hundreds of seconds. It can be seen that the balance between imaging quality and imaging speed is the result of the simultaneous improvement on hardware devices and algorithms. Our deep-learning-based reconstruction algorithm based on the I-sCMOS camera can realize fast imaging at a second-level speed, and while keep high visibility simultaneously. 

In summary, we experimentally demonstrate a fast correlated-photon imaging scheme enhanced by deep learning algorithm to recover objects illuminated with SPDC source. We show the CAE neural network is superior to the classical algorithm in denoising and fast reconstruction at ultra weak illumination. First, the reconstruction process is independent on the prior knowledge, such as Poisson distribution or point spread function. In addition, large networks trained with a large number of images can overcome the uncertainty in the optimization process, where we can find the best solution for the non-convex problem. Furthermore, once trained, the CAE model remain nonlinear and highly complex, which has a good response to the imaging system. Finally, it can complete high-contrast image restoration by collecting a limited number frames of faint picture, which provides the possibility of real-time imaging. In this work, the optimization results are built on single frames, which breaks the rule that it will take more time to collect more photons in order to obtain a high SNR image. Such capabilities are endorsed by the development of hardware and the continuous innovation of new algorithms, paving the way for a broader and practical exploitation of quantum enhanced imaging. 

\subsection*{Acknowledgments.}
The authors thank Jian-Wei Pan for helpful discussions. This work was supported by National Key R\&D Program of China (2019YFA0308700 and 2017YFA0303700); National Natural Science Foundation of China (NSFC) (61734005, 11761141014, 11690033); Science and Technology Commission of Shanghai Municipality (STCSM) (17JC1400403); Shanghai Municipal Education Commission (SMEC) (2017-01-07-00-02-E00049); X.-M. J. acknowledges additional support from a Shanghai talent program.
\\

\subsection*{Data availability.}
The data that support the findings of this study are available from the corresponding author upon reasonable request.
\\

\subsection*{Methods}
\paragraph*{Experimental details:} In the optical experiments, we use a $780nm$ mode-lock Ti:sapphire oscillator with a laser pulse duration of $140fs$ and a repetition rate of $80$ MHz. The laser is frequency doubled to 390nm in a $LiB_3O_5$ (LBO) crystal and then the ultraviolet laser pumps a $2$-mm thick $\beta$-$BaB_2O_4$ (BBO) crystal to create correlated photon pairs via a  type-\uppercase\expandafter{\romannumeral2} SPDC process. The single-channel count rate and two-channel coincidence reach about 500,000 and 100,000, respectively. The amplitude-only SLM contains the maximum number of pixels 1920 $\times$ 1080 and the effective size 8.0 $\mu m$ $\times$ 8.0 $\mu m$. The images size of the handwritten images from the EMNIST and MNIST database is 28 $\times$ 28 pixels. We magnified these images 12 times and displayed them on the SLM. As a result, the physical size of images on the SLM is about 2.34mm $\times$ 2.34mm. The camera is an I-sCMOS camera with a 5.5 megapixel sensor and a 10\% quantum efficiency at $780nm$. 

\paragraph*{CAE model:} The Convolutional Auto-Encoder Model has three convolutional layers, three maxpooling layers, three deconvolutional layers, three unpooling layers as shown in Fig.2(b). The first convolution layer takes a feature map of size ( 28 $\times$ 28 $\times$ 1 ) and outputs a feature map of size ( 28 $\times$ 28 $\times$ 64 ) : it computes 64 filters with size 3 $\times$ 3 over its input.  In the next max-pooling operation, the patches extracted by a 3 $\times$ 3 convolution with stride 2 over a 28 $\times$ 28 input (without padding). The max-pooling operation halves the above feature map to size ( 14 $\times$ 14 $\times$ 64 ). After three convolution and pooling operations, the data has been compressed to a feature map of size ( 4 $\times$ 4 $\times$ 32 ). Decoder is an inverse process similar to encoder. After three upsample and deconvolution operations, we get a feature map of size ( 28 $\times$ 28 $\times$ 64 ). Finally, we reconstruct the optimized images by convolving the above feature map. The activate function of these neurons is Rectified Linear Units (ReLU) which allow for faster and effective training of deep neural architectures on large and complex datasets. The CAE program is implemented using Python version 3.7 and TensorFlow framework, and the implementation is GPU-accelerated with a NVIDIA GTX1650 4G.

\clearpage

\clearpage

\begin{figure}[htbp]
	\centering
	\includegraphics[width=1.0\linewidth]{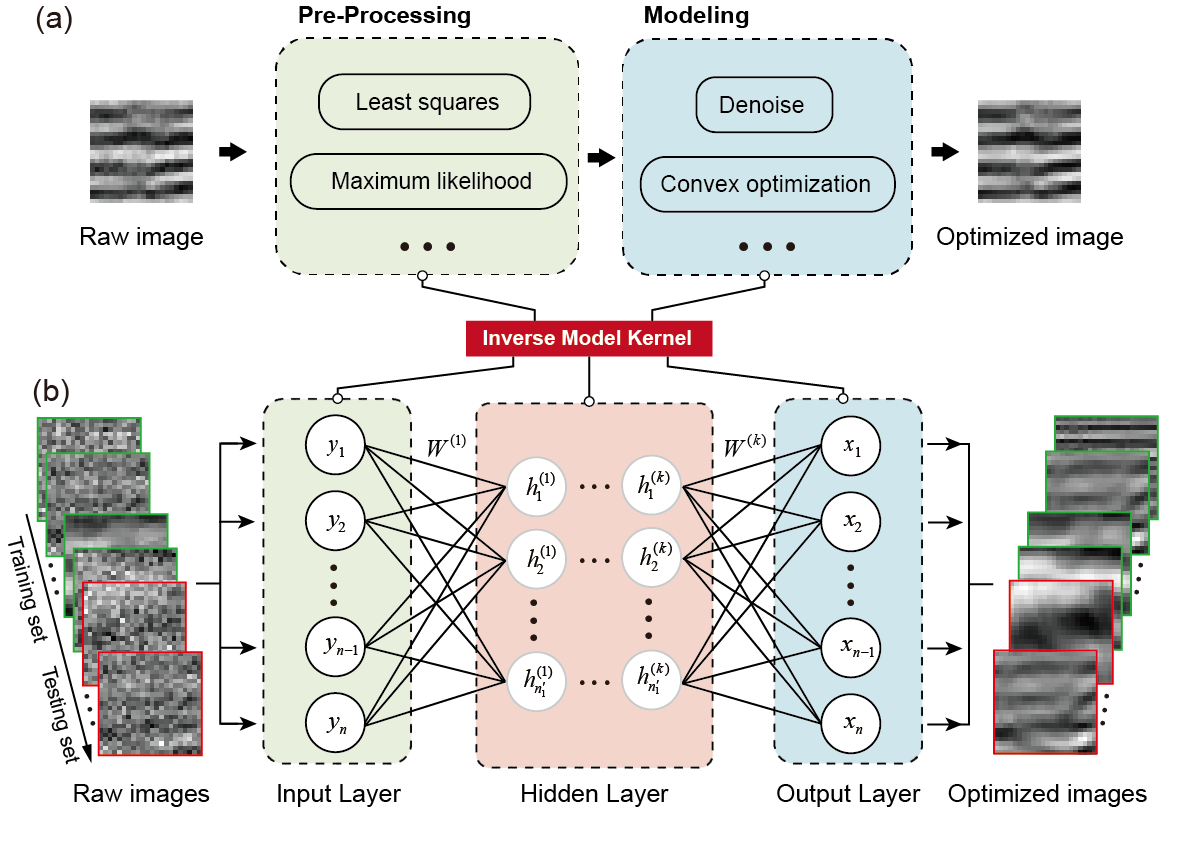}
	\caption{\textbf{Reconstruction algorithms in an inverse imaging problem.} \textbf{(a)} Schematic of the numerical algorithm process. This process only optimizes a single-frame image to achieve a global optimal solution by multiple complex steps such as least squares, maximum likelihood and convex optimization. \textbf{(b)} Schematic of the learning algorithm process. This process builds a network structure to directly connect the input and output images, and then a large set of training images are fed into the network to learn features of the imaging system by optimizing joint parameters. Once training step is finished, we can rebuild optimized images in real time. }
	\label{f1}
\end{figure}

\clearpage

\begin{figure}[htbp]
	\centering
	\includegraphics[width=1.0\linewidth]{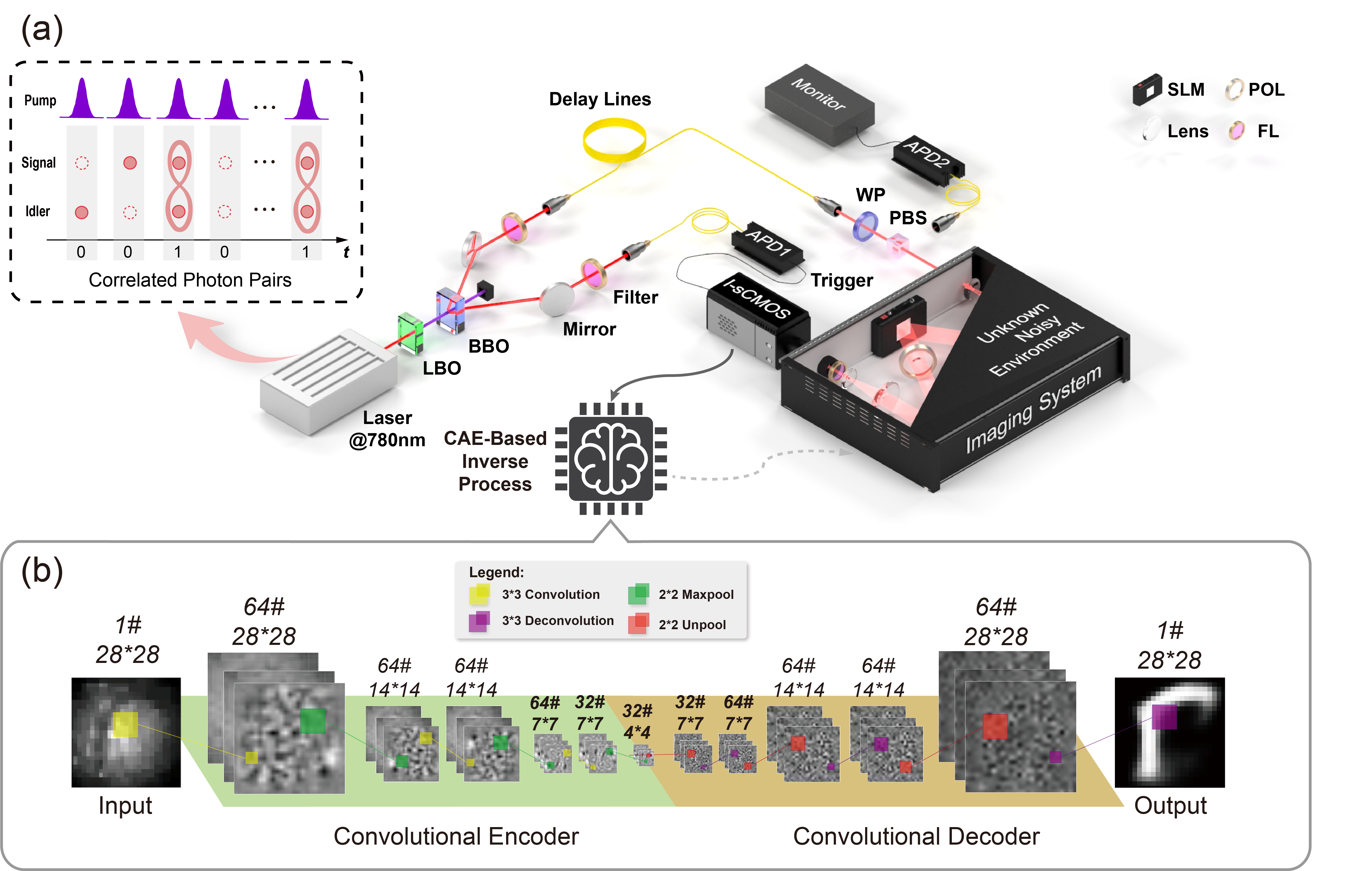}
	\caption{\textbf{Experimental setup and CAE model.}  \textbf{(a)} Sketch of the experimental setup. Correlated photon pairs with a wavelength of 780 nm are simultaneously generated from $\beta$-barium borate (BBO) crystal cut for Type-II phase matching. The signal photons probe an object displayed on a SLM and reflected photons are detected by an I-sCMOS camera, whose intensifier is triggered by a SPAD detector responding to the corrected idler photons. A 10 m fiber delay line is necessary to compensate for the electronic delays between two arms. LBO: $LiB_3O_5$ crystal; WP: wave plate; PBS: polarization beam splitter; APD: avalanched photon diode;  POL: polarizer; FL: filter lens (780$\pm$10nm). \textbf{(b)} CAE model. The structure includes two parts: Encoder and Decoder. The encoder process compresses the input data to lower dimensional representation, while the decoder process decompresses the representation to reconstruct the input as best as possible.}
	\label{f2}
\end{figure}

\clearpage

\begin{figure}[htbp]
	\centering
	\includegraphics[width=1.0\linewidth]{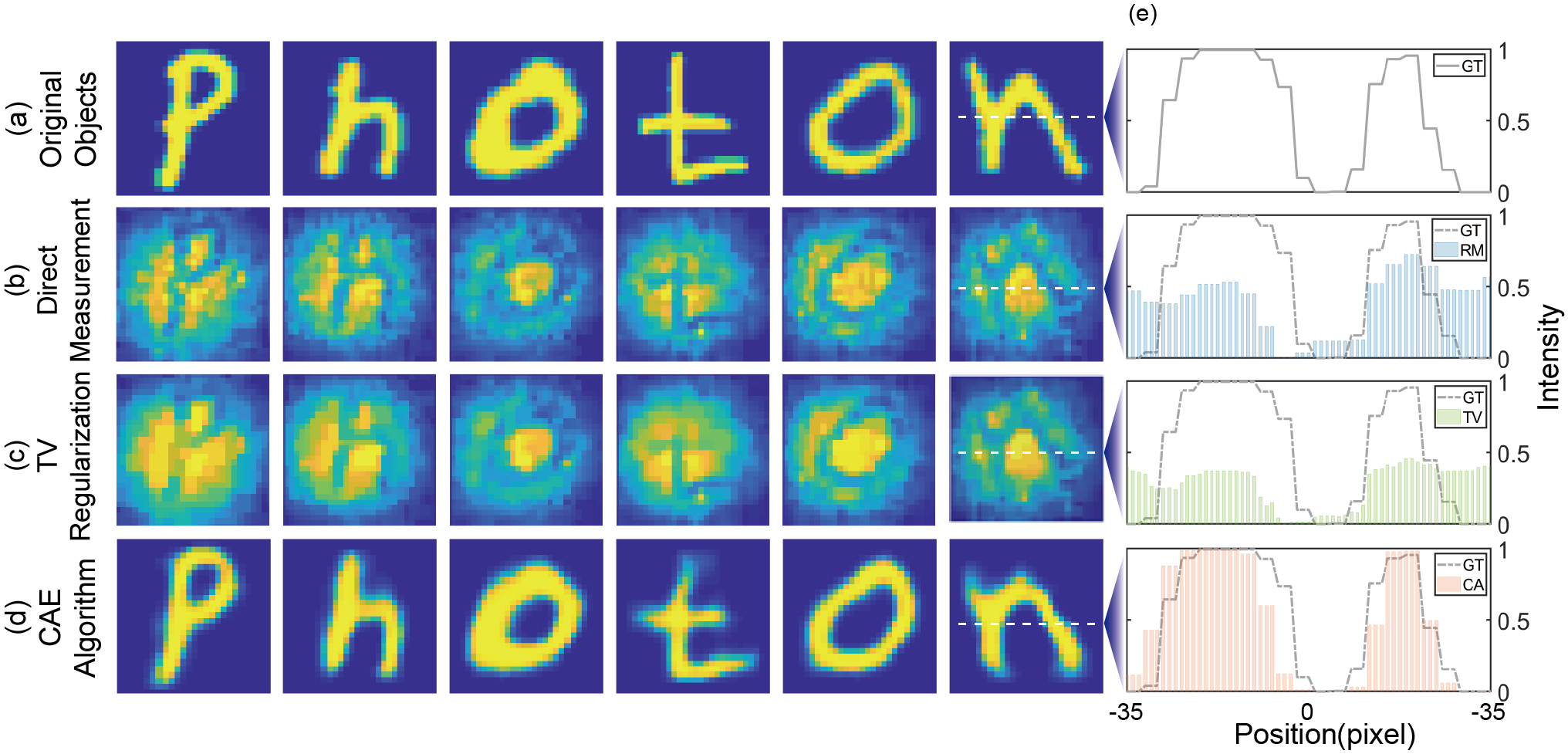}
	\caption{\textbf{Reconstruction results with both numerical and learning algorithms.}  \textbf{(a)} Original object intensity of six letters representing the word ``photon'' from EMNIST dataset.  \textbf{(b)} Direct measurements captured by the I-sCMOS camera with 5s exposure time. \textbf{(c)} TV regularization algorithm reconstruction. \textbf{(d)} CAE algorithm reconstruction. \textbf{(e)} Intensity plots from one line of the ``n'' letter corresponding to the original intensity.}
	\label{f3}
\end{figure}

\clearpage

\begin{figure}[htbp]
	\centering
	\includegraphics[width=1.0\linewidth]{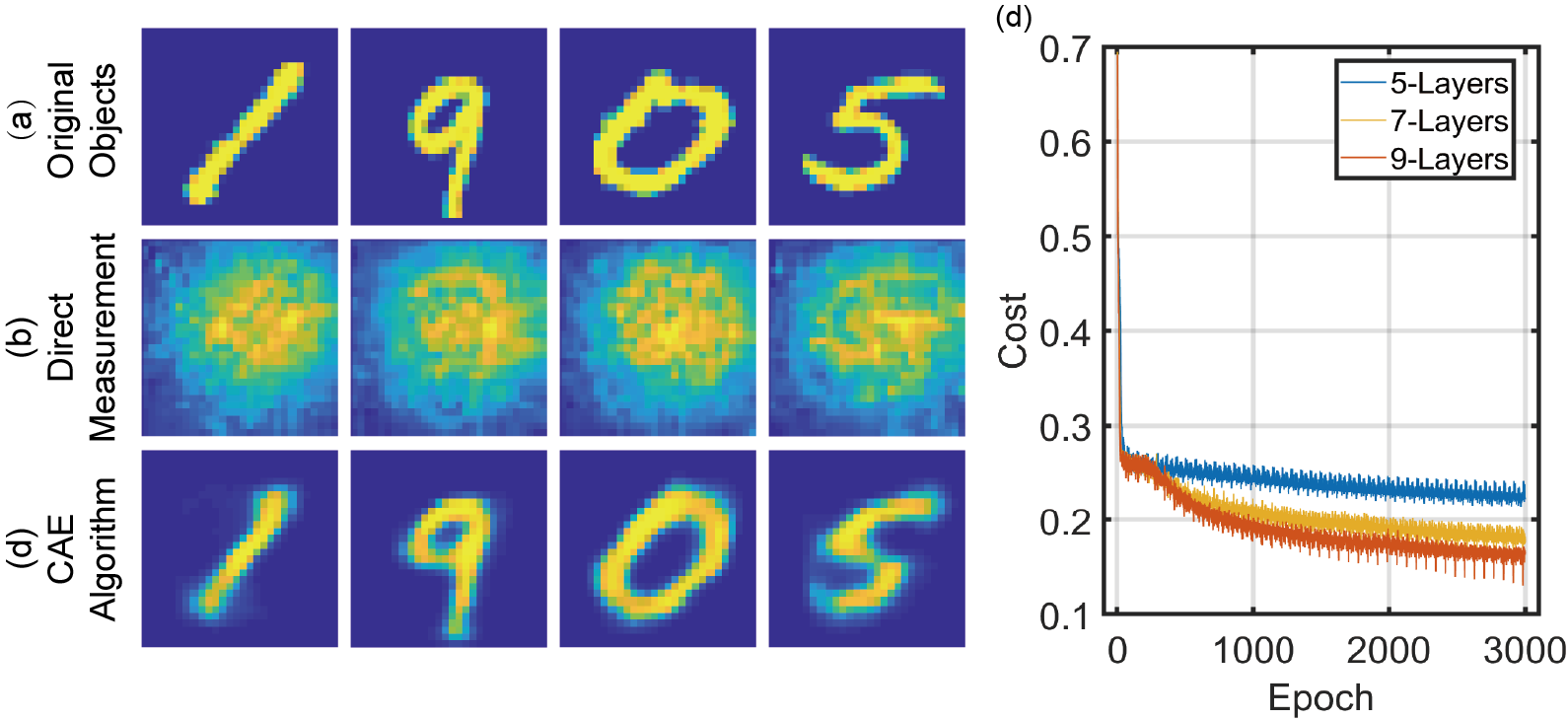}
	\caption{\textbf{Reconstruction results with learning algorithms at 0.8 photons/pixel on average.} \textbf{(a)} Original object intensity of four digits ``1905'' from the test set of MNIST dataset. \textbf{(b)} Direct measurements in the plane of camera.  \textbf{(c)} CAE algorithm reconstructions demonstrate strong robustness in more noisy environment. (d) The mean-squared error (MSE) between the network output and original handwritten digits drops down as epochs increase.}
	\label{f4}
\end{figure}

\clearpage

\begin{figure}[htbp]
	\centering
	\includegraphics[width=1.0\linewidth]{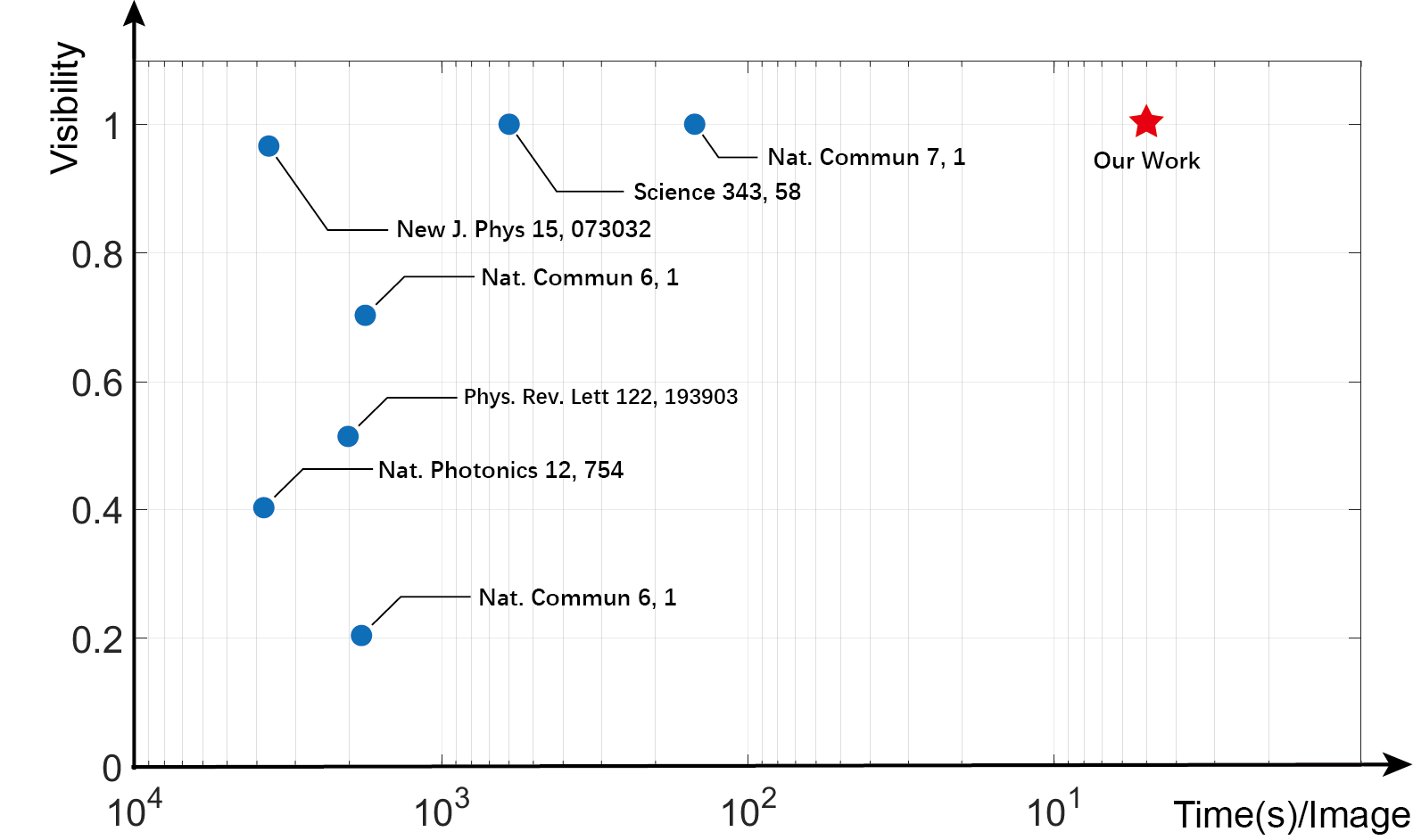}
	\caption{\textbf{Summary of the state-of-the-art imaging experiments at single-photon level.} To achieve high-contrast imaging quality,  numerical algorithms require sparse single photon data per frame, thus intensified CCD and CMOS cameras have to accumulate thousands of frames. The emergence of new imaging devices like SPAD cameras makes less necessary photons and high contrast possible, but numerical algorithms are still a barrier to realize fast imaging. Deep learning algorithms effectively solve this problem, achieving a win-win for both imaging speed and quality. }
	\label{f5}
\end{figure}

\clearpage

\end{document}